\documentclass[12pt]{iopart}

\usepackage{epsfig}

\begin{document}

\title{A supercritical series analysis for the generalized contact process with diffusion}

\author{W.G. Dantas}

\address{Instituto de F\'{\i}sica,
Universidade de S\~{a}o Paulo,
Caixa Postal 66318
05315-970 S\~{a}o Paulo, S\~{a}o Paulo, Brazil}

\author{J.F. Stilck}
\address{Instituto de F\'{\i}sica, Universidade Federal Fluminense,
24210-340, Niter\'oi, RJ, Brazil}

\begin{abstract}
We study a model that 
generalizes the CP with diffusion. An additional transition is
included in the model so that at a particular point of its phase
diagram a crossover from the directed percolation to the compact
directed percolation class will happen. We are particularly interested
in the effect of diffusion on the properties of the crossover between
the universality classes. To address this point, we develop a
supercritical series expansion for the ultimate survival probability
and analyse this series using d-log Pad\'e and partial differential
approximants. We also obtain approximate solutions in the one- and
two-site dynamical mean-field approximations. We find evidences that,
at variance to what happens in mean-field approximations, the
crossover exponent remains close to $\phi=2$ even for quite high
diffusion rates, and therefore the critical line in the neighborhood
of the multicritical point apparently does not reproduce the
mean-field result (which leads to $\phi=0$) as the diffusion rate
grows without bound. 
\end{abstract} 

\pacs{05.70.Ln, 02.50.Ga,64.60.Cn}

\maketitle

\section{Introduction}
\label{intro}
Phase transitions in stochastic models have attracted a great attention
in recent years. Although much work has been done on systems whose
stationary states are described by thermodynamics, also phase
transitions in systems far from equilibrium have been studied in much
detail, and it is remarkable that concepts of scaling and
universality, developed in the context of thermodynamic phase
transitions, have been applied successfully in these new situations as
well. In particular, systems which exhibit absorbing states, and
therefore do not obey detailed balance, are a rich group of models
where out of equilibrium phase transitions have been studied. Much of
this work has been done based on numerical simulations, stochastic
models are very well fitted to these techniques, but analytical
approaches are also useful and may sometimes lead to very precise
results. Therefore, as in other fields of Physics, simulational and
analytical approaches are complementary to study non-equilibrium phase
transitions \cite{md99}.

One of the largest and most studied universality classes in models with
absorbing states is the directed percolation (DP) class, and in this class
the most studied model is the contact process (CP), which in one of
his interpretations was formulated as a very simple model to describe
the evolution of an epidemic disease, which spreads through contact
between healthy and  sick individuals, placed on sites of a
lattice. In this model, at each site of the lattice an individual is
located, and this individual will be in one of two states: healthy or
sick. It is usual to associate a sick individual to a particle and a
healthy one to a hole. Particles are created autocatalytically, that
is, a particle may be created at an empty site with a rate which is
proportional to the number of occupied first neighbor sites of this
site. An occupied site may become empty with a unitary rate
(spontaneous healing). Even in one dimension, a phase transition is
found in the 
model between the absorbing stationary state (in which all individuals
are healthy and no new sick person may be generated by contact, or
else for a lattice without any particle) and a
stationary phase with a nonzero fraction of particles, called
active state \cite{h74}. The contact process is related to several
other stochastic models \cite{s72,gt79,zgb86,tty92}, and the directed
percolation universality class seems to include all models where a
transition between an absorbing and an active state occurs, with a
scalar order parameter, short range interactions and no conservation
laws \cite{j81}. No exceptions to this conjecture have been reported
so far \cite{h00}.

Recently, a generalization of the CP was studied using mean-field
approximations and simulations \cite{dts05}, as well as series
expansions \cite{ds05}. In this model, an additional process is
included besides the autocatalytic creation and the
spontaneous annihilation of particles: the autocatalytic creation
of holes, that is, an occupied site may become empty by two
processes: either spontaneously or with a rate which is proportional
to the number of empty first neighbor sites. It is more natural to
describe this model in terms of two distinct types of particles, $A$
and $B$, requiring that each lattice site at any time is either
occupied by an $A$ or a $B$ particle. The process $B \to A$ occurs
with a rate which is proportional to the number of first neighbor sites
with $A$ particles, whereas the opposite process $A \to B$ may happen
either spontaneously or with a rate proportional to the number of
first neighbor sites occupied by $B$ particles. The CP is a particular
case of this model where no autocatalytic creation of $B$ particles is
allowed. In another particular case, an additional symmetry is present
in the model: when the spontaneous creation of $B$ particles is
suppressed, the model becomes symmetric with respect to the interchange
of $A$ and $B$ particles for equal creation rates of $A$ and $B$
particles, being known 
as the biased voter model 
\cite{bvm}, with a transition between two symmetric absorbing states,
where the lattice is totally filled with the same type of particle. If
the rates of the two processes ($A \to B$ and $B \to A$) are the same,
the density of $A$ and $B$ particles does not change as the system
evolves, but if the rate of one of these processes is larger, the
system will reach the absorbing state in which only the particles
created at a larger rate are present. This model, having an additional
symmetry, belongs to the compact directed percolation (CDP) class, with
critical exponents which are different from the ones found in the
CP. The crossover between both universality classes was the major
motivation to study this generalized model. In particular, the results
of a two-variable series expansion for the ultimate survival
probability of the 
model lead to rather precise evidences that the crossover exponent at
the point where the universality class of the model changes is given
by $\phi=2$, coincident with the mean-field value, which agrees both
with the limits obtained by Liggett \cite{l94} and with simulational
results \cite{tikt95} for the same crossover in the Domany-Kinzel
automaton \cite{dk84}, thus 
providing an evidence that the process of update (parallel in the DK
automaton and sequential in the CP) does not change the value of the
crossover exponent.

It is of interest to study the effect of diffusion on the
behavior of such models. This was already done for the CP, with the
result that the transition point becomes a critical line as the new
variable (rate of diffusion) is introduced in the model
\cite{dj93,mf04}. In this case, as expected, mean-field results are
obtained in the limit where the evolution of the system is dominated
by diffusion, both in the values of the transition rate and of
critical exponents. Thus the limit of infinite diffusion rate may
also be viewed as a crossover transition between non-classical and
classical behavior. These results may be heuristically justified if we
note that in this limit, since diffusion processes are much more
probable that the others, for each other process the local densities
may be replaced by the global ones, eliminating the effect of
fluctuations. 

In this work we include diffusion in the generalized CP model
described above, in order to find out its effect on the phase diagram
and critical exponents of the model. After defining the model more
precisely in section \ref{model}, we obtain its phase diagram in the
two-site cluster mean-field approximation in section \ref{ca}. We then
proceed obtaining supercritical series expansions for the survival
probability in section \ref{se}, which are analysed using Pad\'e
approximants and two-variable partial differential approximants in
section \ref{as}. Conclusions and final comments may be found in
section \ref{conc}.

\section{Definition of the model}
\label{model}

The model is defined on a one-dimensional lattice with $N$ sites and periodic
boundary conditions. Each site is occupied either by a particle $A$ or a
particle $B$, no holes are allowed. The microscopic state of the model may thus
be described by the set of binary variables
$\eta=(\eta_1,\eta_2,\ldots,\eta_N)$, where $\eta_i=0$ or $1$ if site $i$ is
occupied by particles $B$ or $A$, respectively. 

The model evolves in time according to the following Markovian rules:
\begin{enumerate}
\item A site $i$ of the lattice is chosen at random.
\item If the site is occupied by a particle B, it becomes occupied by a
  particle A with a transition rate equal to $p_a n_A/2$, where $n_A$ is the
  number of A particles in the sites which are first neighbors to site $i$.
  
  $$\framebox{A}\mbox{ }B\mbox{ }\framebox{B}\stackrel{\tiny{p_a/2}}
  {\longrightarrow}\framebox{A}\mbox{ }A\mbox{ }\framebox{B}$$
  
\item If site $i$ is occupied by a particle A, it may become occupied by a
  particle B through three processes:
  \begin{itemize}
  \item Spontaneously, with a transition rate $p_c$.
    $$A\stackrel{\tiny{p_c}}{\longrightarrow}B$$
  \item Through an autocatalytic reaction, with a rate $p_bn_B/2$, where $n_B$
    is the number of B particles in the sites which are first neighbors to site
    $i$. 
    $$\framebox{A}\mbox{ }A\mbox{ }\framebox{B}\stackrel{\tiny{p_b/2}}
    {\longrightarrow}\framebox{A}\mbox{ }B\mbox{ }\framebox{B}$$
  \item By interchanging the $B$ particle with an $A$ particle located
    in a first neighbor site to site $i$. This transition occurs with
    rate equal $\tilde{D}$. 
    $${ }A\mbox{ }\framebox{B}\stackrel{\tiny{\tilde{D}}}
    {\longrightarrow}B\mbox{ }\framebox{A}$$
  \end{itemize}
  where $\framebox{\mbox{}}$ indicates the state of the first neighbour.
\end{enumerate}

We define the time in such a way the the non-negative rates $p_a$, $p_b$,
and $p_c$ obey the normalization $p_a+p_b+p_c=1$. We may then discuss the
behavior of the model in the $(p_a,p_c,\tilde{D})$ space without loss of 
generality.

When $p_c\neq 0$ the configuration where all the sites are 
occupied by particles $B$ is absorbing, while for $p_c=0$ both 
symmetrical configurations in which all sites are occupied by the same
particles are absorbing, independently of the value of $\tilde{D}$. In
the particular case $p_b=0$, in which the model corresponds to the CP
with diffusion, there exists a critical line in the
$(p_a,p_b,\tilde{D})$ space where an active configuration (with
nonzero density of $A$ particles) becomes identical with the absorbing
one. This line becomes a critical surface in the general case $p_b
\neq 0$.  Based on the symmetries of the model this surface must be of
second order  
with critical exponents of DP class for finite values of the diffusion
rate $\tilde{D}$, except for the case $p_c=0$. 
In this case, the transition occurs between two 
absorbing states (lattice full by particles $A$ or $B$) at the 
the point $p_a=1/2$, $\forall\tilde{D}$, with critical 
exponents of the CDP class. 

For a certain fixed value of $\tilde{D}$, the behavior of any
stationary density close to multicritical point $(p_a=1/2,p_c=0)$
should exhibit the scaling form,

\begin{eqnarray}
\label{eq1}
g(p_a-1/2,p_c,\tilde{D})\sim
(p_a-1/2)^{e_g(\tilde{D})}F\left(\frac{p_c}{|p_a-1/2|^{\phi(\tilde{D})}}\right). 
\end{eqnarray}

The critical exponent associated with the density variable $g$,
$e_g(\tilde{D})$, should correspond to the CDP universality class, and the
scaling function $F(z)$ is singular at a value $z_0(\tilde{D})$ of its argument,
which corresponds to the critical line for a given value of $\tilde{D}$.
Thus, the critical line is asymptotically given by 
$p_c =z_0(\tilde{D})(pa-1/2)^{\phi(\tilde{D})}$. In this way we have two
exponents as a function 
of the diffusion and it is interesting to find out how the exponents
change as diffusion processes are introduced in the model.

\section{Cluster Approximations}
\label{ca}

We have derived solutions for the cluster dynamic approximations in
simple mean-field  
(one-site) and two-site levels \cite{md99}. The simple mean-field
solution is independent of the diffusion rate, and the critical line
is located at
$p_a=1/2$, independently of $p_c$. Thus, in this approximation the
crossover exponent vanishes identically. The lowest order of
mean-field cluster approximation in which the effect of diffusion is
present in the results is the two-site approximation. Without going
into the details of this approximation, since the calculations are
similar to the ones performed recently in the model without diffusion
\cite{dts05}, in the two-site level cluster approximation the critical
line is given by:

\begin{eqnarray}
\label{eq2}
p_a^c=\frac{1}{8}\left[4-(p_c+2\tilde{D})+
  \sqrt{(p_c+2\tilde{D})^2+8p_c}\right]. 
\end{eqnarray}

We notice that as the value of  $\tilde{D}$ grows, the critical line
approaches the result obtained in the one-site approximation, as
expected. If we 
expand equation \ref{eq2} for small values of $p_c$, we obtain
$p_a^c-1/2 \approx 2p_c/\tilde{D}$, showing that the two-site
approximation leads to the crossover exponent $\phi=1$ for any finite
value of the diffusion rate $\tilde{D}$ and $\phi=0$ in the limit of
infinite diffusion rate. For vanishing diffusion rate, we obtain the
crossover exponent $\phi=2$. 

\begin{figure}[h!]
\begin{center}
\vspace*{0.5cm}
\epsfig{file=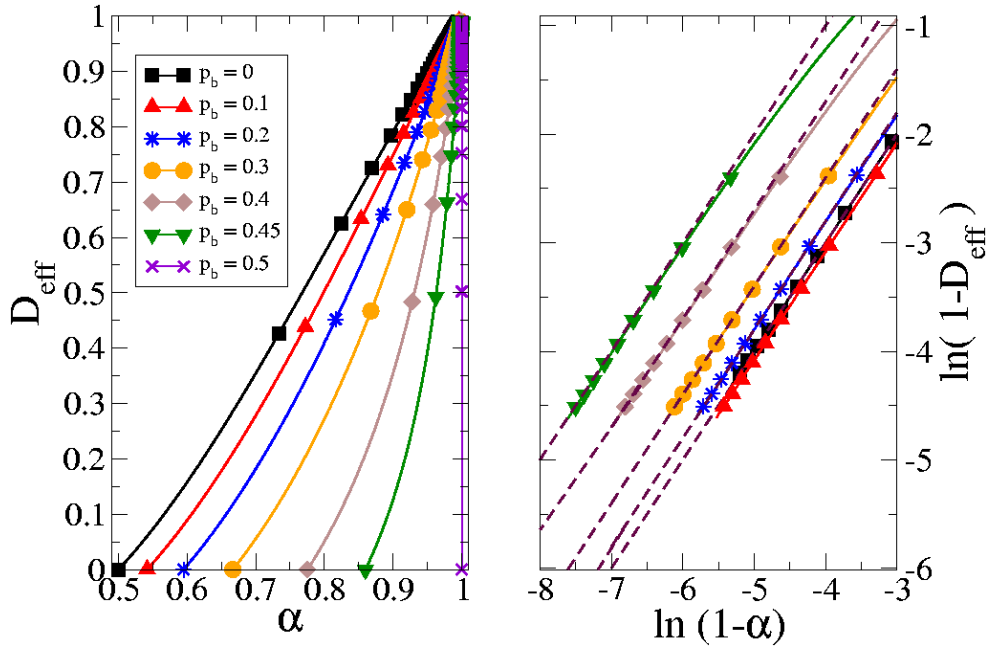,scale=0.5}
\vspace{0.4cm}
\caption{Results of the two-site approximation. In the
  left panel we see different phase diagrams for fixed values 
  of $p_b$ using the new variables $\alpha$ and $D_{eff}$. In the
  right we have the estimates of the crossover exponent in each
  of these diagrams. In all cases the parallel dashed lines indicate
  that this exponent is $\tilde{\phi}=1$ for any value of the
  parameter $p_b<1/2$.} 
\label{fig2}
\end{center}
\end{figure}

Another point which is worth noticing is the crossover as the
diffusion rate approaches infinity, where mean-field behavior is
expected. For this purpose, we may define the variables
$\alpha=(1-p_a)/p_a$ and $D_{eff}=\alpha\tilde{D}/
(1+\alpha\tilde{D})$. Rewriting the expression for the critical
condition \ref{eq2} in terms of the variables $x=1-\alpha$,
$y=1-D_{eff}$, and $p_b$, we obtain:
\begin{equation}
y=\frac{x(x-2)}{p_bx^3+(1-4p_b)x^2+(5p_b-1)x+1-2p_b}.
\end{equation}
It is apparent that, as long as $p_b<1/2$, we have
$\tilde{\phi}=1$ for the crossover between the regimes of finite and
infinite diffusion. In the particular case $p_b=0$, which corresponds
to the CP, we recover two-site approximation results obtained before
\cite{mf04}. For the particular case $p_b=1/2$, which corresponds to
the CDP limit, we have the solution $x=0$ for any value of $y$, so
that the locus of the transition is not affected by the
diffusion. These results are illustrated in figure \ref{fig2}.

\section{Derivation of the supercritical series for the model}
\label{se}

Now let us develop a three-variable supercritical series expansion for the
model. We follow closely the operator formalism presented in the paper by
Jensen and Dickman on series for the CP process and related models
\cite{dj91}. We may represent the microscopic configurations of the
lattice by the direct product of kets
\begin{equation}
| \eta \rangle= \bigotimes_i | \eta_i \rangle,
\end{equation}
which are defined to be orthonormal
\begin{equation}
\langle \eta | \eta^\prime \rangle = \prod_i \delta_{\eta_i,\eta_i\prime}.
\end{equation}
Now we may define A particles creation and annihilation operators for the site
$i$: 
\begin{eqnarray}
A_i^\dagger |\eta_i \rangle &=&(1-\eta_i)|\eta_i+1\rangle,\nonumber \\
A_i |\eta_i\rangle &=& \eta_i|\eta_i-1\rangle.
\end{eqnarray}
In this formalism, the state of the system at time $t$ may be
represented as
\begin{equation}
|\psi(t) \rangle = \sum_{\{\eta\}} p(\eta,t) |\eta \rangle.
\end{equation}
If we define the projection onto all possible states as
\begin{equation}
\langle \; | \equiv \sum_{\{\eta\}} \langle\eta|,
\end{equation}
the normalization of the state of the system may be expressed as
$\langle \; 
|\psi \rangle =1$. In this notation, the master equation for the
evolution of the state of the system is
\begin{equation}
\frac{d |\psi(t)\rangle}{d t}=S|\psi(t)\rangle.
\label{me1}
\end{equation}
The evolution operator $S$ may be expressed in terms of the creation and
annihilation operators as $S=\lambda R + V$ where
\begin{eqnarray}
R&=& \sum_i [\gamma(2-A_{i-1}^\dagger A_{i-1} +
A_{i+1}^\dagger A_{i+1})+1](A_i-A_i^\dagger A_i)+\nonumber\\
&+&\bar{D}\sum_i (1-A_{i-1}^{\dagger}A_i)A_{i-1}A_i^{\dagger}+
(1-A_{i+1}^{\dagger}A_i) 
A_{i+1}A_i^{\dagger}, \\
V &=& \sum_i (A_i^\dagger + A_i^\dagger A_i -1)(A_{i-1}^\dagger A_{i-1} +
A_{i+1}^\dagger A_{i+1}),
\end{eqnarray}
and the new parameters 
\[
\lambda \equiv \frac{2p_c}{p_a},
\]
\[
\gamma \equiv \frac{p_b}{2p_c}
\]
and
\[
\bar{D} \equiv \frac{\tilde{D}}{\lambda}
\]
were introduced.

We notice that the operator $R$ includes the diffusion and the
annihilation of A particles processes, while the creation of A
particles is present in the operator $V$. Therefore, at small values
of the parameter $\lambda$ creation of A particles is favored, and the
decomposition above is convenient for a supercritical perturbation
expansion. Let us show explicitly the effect
of each operator on a generic configuration $(\mathcal{C})$. 
\begin{eqnarray}
R({\mathcal{C}})&=&\gamma \sum_i
({\mathcal{C}}^\prime_i)+2\gamma \sum_j
({\mathcal{C}}^\prime_j) + \sum_k
({\mathcal{C}}^\prime_k)+\nonumber\\
&+&\bar{D}\left[\sum_r({\mathcal{C}}^{\prime\prime\prime}_r)+ 
\sum_s({\mathcal{C}}^{\prime\prime\prime r}_s)+
\sum_t({\mathcal{C}}^{\prime\prime\prime l}_t)\right]+\nonumber\\
&-&[\gamma(r_1+2r_2)+\bar{D}(t_1+2t_2)+r]({\mathcal{C}}),
\label{as0}
\end{eqnarray}
where the first sum is over the $r_1$ sites with A particles and one B
neighbor, 
the second sum is over the $r_2$ sites with A particles and two B neighbors,
the third 
sum is over all $r$ sites with A particles of the configuration
$(\mathcal{C})$. The next sum is again over the $r_1$ sites with A
particles  
and one B neighbor and the two last sums are over the $r_2$ sites with 
A particles and two B neighbors. 
Configuration $({\mathcal{C}}_i^{\prime})$ is obtained  replacing
the A particle at site $i$ by a B particle,
$({\mathcal{C}}^{\prime\prime\prime}_i)$ 
is obtained interchanging the A particle at the site $i$ with its
single B neighbor and finally,
$({\mathcal{C}}^{\prime\prime\prime (r,l)}_i)$ is a configuration
where the A particle 
at the site $i$ is interchanged with the B particle located at the
right ($r$) or to the left ($l$) of site $i$.
It is convenient associate the diffusion with the annihilation process 
to avoid some ambiguity in truncating the series in a certain order.
In the other hand, the action of operator $V$ is
\begin{equation}
V({\mathcal{C}})=\sum_i ({\mathcal{C}}^{\prime \prime}_i)+2 \sum_j
({\mathcal{C}}^{\prime \prime}_j) -(q_1+2q_2)({\mathcal{C}}),
\label{av}
\end{equation}
where the first sum is over the $q_1$ sites with B particles and one A
neighbor, 
the second sum is over the $q_2$ sites with B particles and two A
neighbors. Configuration $({\mathcal{C}}_i^{\prime \prime})$ is obtained
replacing the B particle at site $i$ in configuration $(\mathcal{C})$ by a A
particle.

Since we are interested in the long time behavior of the system, it is
useful to consider Laplace transforms. For the state ket we have:
\begin{equation}
|\tilde{\psi}(s) \rangle = \int_0^\infty e^{-st} |\psi(t)\rangle,
\label{lt}
\end{equation}
and inserting the formal solution $|\psi(t)\rangle =e^{St} |\psi(0)\rangle$ of
the master equation \ref{me1}, we find
\begin{equation}
|\tilde{\psi}(s) \rangle = (s-S)^{-1} |\psi(0)\rangle.
\label{tpsi}
\end{equation}
The stationary state $|\psi(\infty) \rangle \equiv \lim_{t \to \infty} |\psi(t)
\rangle$ may then be found noticing that 
\begin{equation}
|\psi(\infty) \rangle = \lim_{s \to 0} s |\tilde{\psi}(s) \rangle,
\end{equation}
which may be obtained integrating \ref{lt} by parts. A perturbative expansion
may be obtained assuming that $|\tilde{\psi}(s) \rangle$ may be expanded in
powers of $\lambda$ and using \ref{tpsi},
\begin{equation}
|\tilde{\psi}(s) \rangle = |\tilde{\psi}_0 \rangle+\lambda |\tilde{\psi}_1
\rangle +\lambda^2 |\tilde{\psi}_2 \rangle + \cdots = \frac{1}{s- V -\lambda
R } |\psi(0) \rangle. 
\end{equation}
Since
\begin{equation}
\frac{1}{s- V -\lambda R }= \frac{1}{s-V} \left[ 1 + \lambda \frac{1}{s-V}
R 
+ \lambda^2 \frac{1}{(s-V)^2} R^2 + \cdots \right],
\end{equation}
we arrive at
\begin{eqnarray}
|\tilde{\psi}_0 \rangle &=& \frac{1}{s-V} |\psi(0)\rangle \nonumber \\
|\tilde{\psi}_1 \rangle &=& \frac{1}{s-V} R |\tilde{\psi}_0 \rangle
\label{rr} \\
|\tilde{\psi}_2 \rangle &=& \frac{1}{s-V} R |\tilde{\psi}_1 \rangle
\nonumber \\
 &\vdots& 
\end{eqnarray}
The action of the operator $(s-V)^{-1}$ on an arbitrary configuration
$({\mathcal{C}})$ may be found noting that
\begin{equation}
(s-V)^{-1} ({\mathcal{C}})=s^{-1} ({\mathcal{C}})+ \frac{V}{s(s-V)}
({\mathcal{C}}), 
\end{equation}
and using the expression \ref{av} for the action of the operator $V$, we get
\begin{equation}
(s-V)^{-1} ({\mathcal{C}})= s_q \left\{ ({\mathcal{C}}) + (s-V)^{-1} \left[
\sum_i 
({\mathcal{C}}^{\prime \prime}_i)+2 \sum_j ({\mathcal{C}}^{\prime \prime}_j)
\right] \right\},
\label{sv}
\end{equation}
where the first sum is over the $q_1$ sites with B particles and one A
neighbor, the second sum is over the $q_2$ sites with B particles and two A
neighbors, and we define $s_q \equiv 1/(s+q)$, where $q=q_1+2 q_2$. 

It is convenient to adopt as the initial configuration a translational
invariant one with a single A particle (periodic boundary conditions are
chosen). Now we may notice in the recursive expression \ref{sv} that the
operator $(s-V)^{-1}$ acting on any configuration generates an infinite set of
configurations, and thus we are unable to calculate $|\tilde{\psi}\rangle$ in
a closed form. We may, however, calculate the extinction probability
$\tilde{p}(s)$, which corresponds to the coefficient of the vacuum state
$|0\rangle$. As happens also for the models related to the CP studied in
\cite{dj91} configurations with more than $j$ particles only
contribute at 
orders higher than $j$, and since we are interested in the ultimate survival
probability for A particles $P_\infty=1-\lim_{s \to 0} s \tilde{p}(s)$, $s_q$
may be replaced by $1/q$ in Eq. \ref{sv}. An illustration of this procedure 
may be seen in \cite{ds05}.

The algebraic operations above may be easily performed in a computer using a
proper algorithm. The
configurations are expressed as binary numbers and the coefficients as double
precision variables. 
With rather modest computational resources (Athlon MP2200, double
processor, 1Gb 
memory) it is not difficult to calculate the coefficients up to order 22. The
required processing time amounts to about 2 hours, the limiting factor is
actually the memory required for the calculation. We define the
coefficients $b_{i,j,k}$ as:

\begin{eqnarray}
\label{pinf}
P_{\infty}=1-\frac{1}{2}(2\gamma+1)\lambda-\frac{1}{4}(2\gamma+1)\lambda^2-
\sum_{i=3}^{22}\sum_{j=0}^{i-1}
\sum_{k=0}^{i-2}b_{i,j,k}\lambda^i\gamma^j\bar{D}^k.
\end{eqnarray}
The case $\alpha=0$ is leads to a series which is
identical the the one for the CP with diffusion obtained in
\cite{dj93}. Also, for $\bar{D}=0$ the two-variable series for the
model without diffusion is \cite{ds05} is recovered. The set of
coefficients is too large to be included here, but may be obtained
upon request from the authors.

\section{Analysis of the series}
\label{as}

To obtain estimates of the critical properties of this model
from the supercritical series for the ultimate survival probability,
given by equation \ref{pinf}, we start by using
the d-log Padé approximants approach. These approximants
are defined as ratios of two polynomials 

\begin{eqnarray}
F_{LM}(\lambda)=\frac{P_L(\lambda)}{Q_M(\lambda)}=
\frac{\sum_{i=0}^{L}p_i\lambda^i} 
{1+\sum_{j=1}^M q_j\lambda^j}=f(\lambda).
\end{eqnarray}
In our case the function $f(\lambda)$ represents the series for
$\frac{d}{d\lambda} 
\ln P_{\infty}(\lambda)$, where we fix the two remaining parameters
$\gamma$ and $\bar{D}$. Therefore, we may obtain approximants with
$L+M\leq 22$, and since it is known that diagonal ($L=M$) and
near-diagonal approximants usually exhibit better convergence
properties, we restrict our calculations to the set of approximants
such that $L=M+\theta$, with $\theta=0,\pm 1$. We built approximants
for the values of $\gamma$ ranging between 0 and 45,
using $D=\bar{D}/(1+\bar{D})=0,0.1,0.2,0.3,0.4,0.5,0.7,0.8$ and
$0.85$. In the table \ref{coefs} some estimates for the critical value
of the parameter $\lambda$ are shown. The estimates poles are
roots of the polynomial $Q_M$ (poles of the approximant).

\begin{table}
\begin{center}
\begin{tabular}{ccccc}
\hline
D&$\gamma$&L&M&$\lambda_c$\\
\hline
0  &0.2&10&10&0.49472\\
   &   &11&11&0.49472\\
0.1&0.2&10&10&0.50476\\
   &   &11&11&0.50476\\
0.7&0.2&10&10&0.64708\\
   &   &11&11&0.64750\\
0.8&0.2&10&10&0.75146\\
   &   &11&11&0.71941\\
\hline
\end{tabular}
\caption{Examples of estimates for critical points obtained by d-log Padé
  approximants. Notice that as the value of $D=\bar{D}/(1+\bar{D})$
  grows the dispersion in the 
  estimates obtained from different approximants for this critical point
  also grows.} 
\label{coefs}
\end{center}
\end{table}

\begin{figure}[h!]
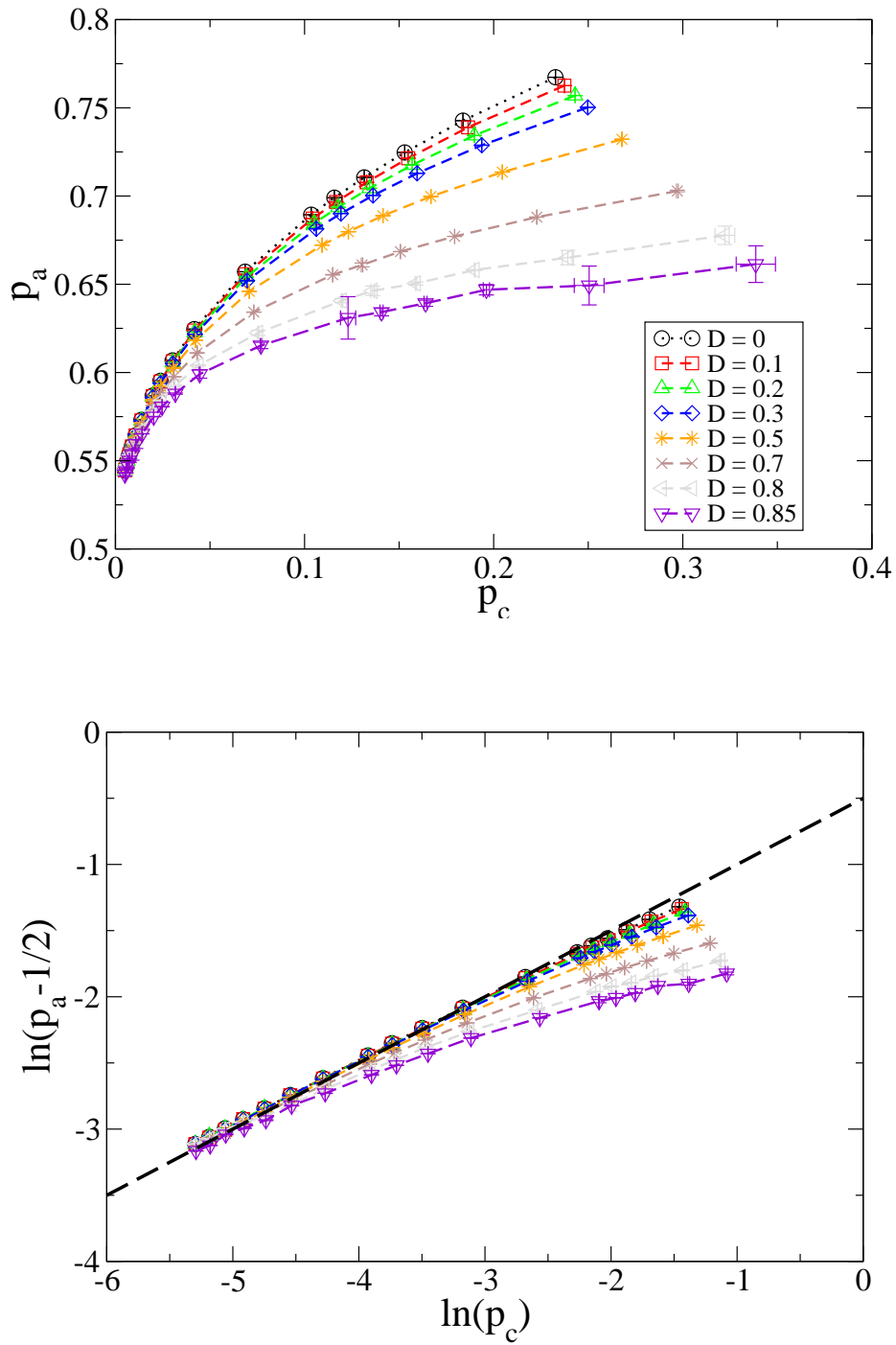

\begin{center}
\vspace*{0.5cm}
\epsfig{file=./linhascrits.eps,scale=0.5}\\
\vspace*{1.3cm}
\epsfig{file=./cross.eps,scale=0.5}
\vspace{0.4cm}
\caption{It the top panel we see the phase diagrams of the model
  obtained from d-log Padé approximants 
  for several values of the diffusion rate $D=\bar{D}/(1+\bar{D})$. The
  graph in the bottom shows log-log plot of the same curves 
  with the dashed line indicating the value of $\phi^{-1}=1/2$.}
\label{fig3}
\end{center}
\end{figure}

With the data provided by the set of approximants, we obtained
the critical lines using the original 
variables $p_a$ and $p_c$, for different values of the diffusion rate
$\bar{D}$.   
Each point of these curves, shown in the figure \ref{fig3}, was
calculated as the average of the estimates provided by the set of
approximants, and the error bar associated to it corresponds to the
standard deviation of the estimates. As the diffusion rate increases,
the dispersion of the approximants also grows and the precision of 
the result becomes smaller generating relatively higher errors.  
Even so, it is possible to notice that all the  
critical curves approach of the multicritical point with a quadratic
curvature, and thus the present results are different from the ones
provided in the mean-field two-sites approximation shown above, which
lead to $\phi=1$.

\begin{figure}[h!]
\begin{center}
\vspace*{1.0cm}
\epsfig{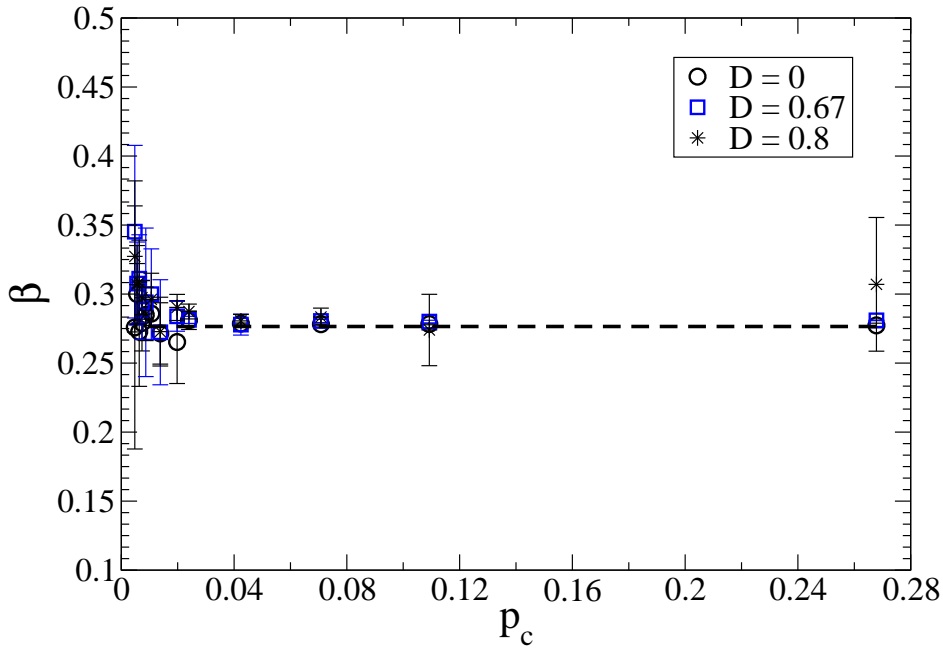}
\vspace{0.3cm}
\caption{Behavior of the exponent $\beta$ as a function of the parameter
  $p_c$ for three different values of the diffusion rate. The value
  corresponding to the DP universality class is shown as a dashed line.}
\label{fig4}
\end{center}  
\end{figure}

The critical exponent $\beta$ associated to the order parameter also can be
estimated from
the d-log Padé approximants, calculating the residue associated to the
physical pole of  
each approximant.  In the figure \ref{fig4}, the average values of the 
estimates for $\beta$ are depicted for different values of the
diffusion rate. They are consistent with the conjecture that the value
of the DP universality class $\beta_{DP}=0.276486(8)$ \cite{jens99} 
applies to the model with finite
rates of diffusion. Also, as expected, we notice that the dispersion of
the estimates grows as the CDP point is approached ($p_c \to 0$)

The behavior of the model in the limit of infinite diffusion rate is
out of the reach  
of this analysis, partially because the reduction of the series to a variable 
leads to poor results in the neighborhood of a multicritical point, as is known 
of other similar cases \cite{fk77} and therefore the
d-log Padé approximants suffer great fluctuations in the region of
high diffusion, as already found in the model without diffusion \cite{ds05}.  

In an attempt to improve the series analysis we may use the Partial
Differential Approximants (PDA' s), which are appropriate for the
study of two-variables series  
of models which undergo multicritical phenomena \cite{fk77}. They may
be regarded as 
a generalization to two variables of the d-log Pad\'e approximants. The defining
equation of the approximants is 
\begin{equation}
P_{\mathbf L}(x,y)F(x,y)=Q_{\mathbf M}(x,y)\frac{\partial F(x,y)}{\partial x} +
R_{\mathbf N}(x,y) \frac{\partial F(x,y)}{\partial y},
\label{pda}
\end{equation}
where $P$, $Q$, and $R$ are polynomials in the variables $x$ and $y$ with the
set of nonzero coefficients ${\mathbf L}$, ${\mathbf M}$, and ${\mathbf N}$,
respectively. The coefficients 
of the polynomials are obtained through substitution of the series expansion
for the quantity which is going to be analyzed 
\begin{equation}
f(x,y)=\sum_{k,k^\prime=0} f(k,k^\prime)x^k y^{k^\prime}
\end{equation}
into the defining equation \ref{pda} and requiring the equality to hold for a
set of indexes defined as ${\mathbf K}$. The coefficients of the
polynomials are then found solving a set of linear equations. Since the
coefficients  
$f_{k,k^\prime}$ of the series are known for a finite set of indexes this sets
an upper limit to the number of coefficients in the polynomials. The
number of equations has to match the number of unknown coefficients,
and thus $K=L+M+N-1$ (as in the Pad\'e approximants, one
coefficient is fixed arbitrarily). An additional issue, 
which is not present in the one-variable case, is the symmetry of the
polynomials. Two frequently used options are the triangular and the
rectangular arrays of coefficients. The choice of these symmetries is related
to the symmetry of the series itself \cite{s90}.

Let us suppose that the quantity represented by the series is expected
to have a multicritical behavior at a point $(x_c,y_c)$, described by
the scaling behavior:
\begin{equation}
f(x,y) \approx |\Delta \widetilde{x}|^{-e_f}Z\left( \frac{|\Delta
\widetilde{y}|}{|\Delta \widetilde{x}|^\phi} \right),
\label{mcs}
\end{equation}
where
\begin{equation}
\Delta \widetilde{x}=(x-x_c )-(y-y_c)/e_2,
\end{equation}
and
\begin{equation}
\Delta \widetilde{y}=(y-y_c)-e_1(x-x_c).
\end{equation}
Here $e_f$ is the critical exponent of the quantity described by $f$ when
$\Delta \widetilde{y}=0$, $e_1$ and $e_2$ are the scaling slopes \cite{fk77}
and 
$\phi$ is the crossover exponent. The function $Z(z)$ is singular for one or
more values of its argument, corresponding to the critical line(s) incident on
the multicritical point. Once the coefficients of the defining polynomials are
obtained, the estimated location of the multicritical point corresponds to the
common zero of the polynomials $Q_M$ and $R_N$. This may be seen substituting
the scaling form \ref{mcs} in the defining equation \ref{pda} of the
approximant. The exponents and scaling
slopes may also be obtained directly from the polynomials, without
integrating the partial differential equation. A detailed discussion of the
algorithm, as well as computer codes, may be found in \cite{s90}.

\begin{figure}[h!]
\begin{center}
\vspace*{1.1cm}
\epsfig{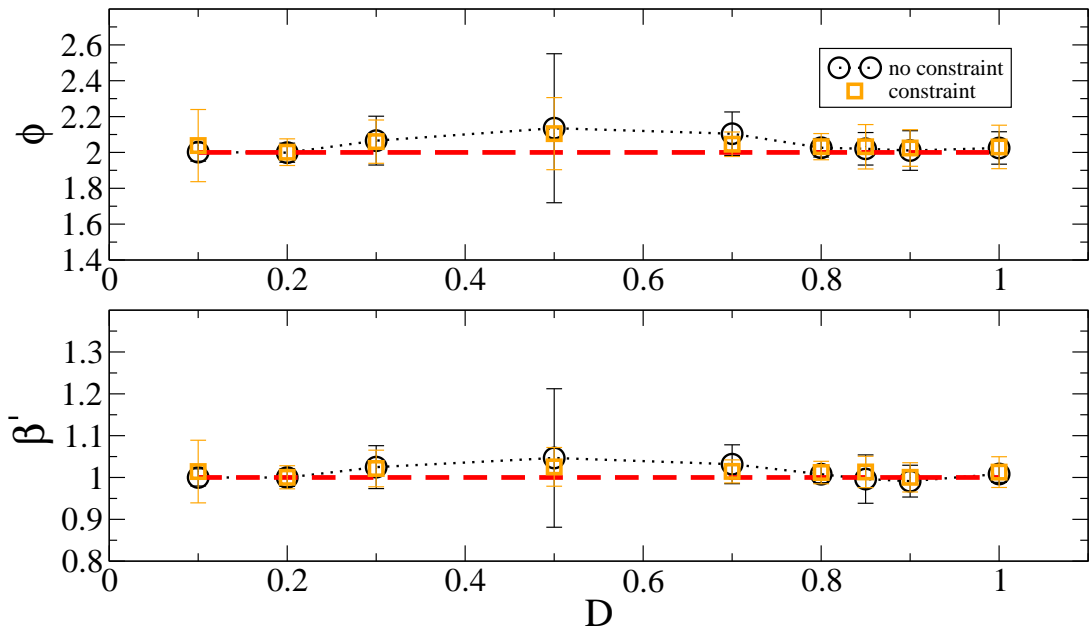}
\vspace{0.4cm}
\caption{Behavior of the exponents $\phi$ and $\beta^\prime$ as a
  function of the diffusion rate. 
  The circles represents the non-constrained results and the squares
  denote the constrained results. The dotted line is only a guide for
  the eyes, while the dashed  
  lines correspond to the values $\phi=2$ and $\beta'=1$ of the case
  without diffusion.} 
\label{fig5}
\end{center}  
\end{figure}

Since in our original series we have three variables, to accomplish
the analysis using PDA's, we fix the value of the variable $\bar{D}$   
and we generate series in the following the variables $x=\lambda$ and 
$y=\gamma\lambda$, to avoid numerical errors due to the divergence
of the variable $\gamma$ in the multicritical point of crossover to
the CDP universality class.  
With that, for each value of $\bar{D}$ we calculate
about 22 approximants, choosing different sizes and 
configurations for the set $\{K,L,M,N\}$, but maintaining 
the constraint $\bf{K}\approx\bf{M}$. The results,
calculated as averages over the estimates of the set of approximants, 
for the exponents $e_f=\beta^{\prime}$ and $\phi$ as functions of the
diffusion rate $D$ 
are shown in the figure \ref{fig5}. The same quantities were
calculated imposing as a constraints the location of the multicritical point, 
fixed at $x=0$ and $y=1$.  The values of the exponents with this constraint 
are represented as squares of the figure \ref{fig5} and are very close
to the 
non-constrained estimates (represented by circles). Again the error
bars are estimated from the dispersion of the results in the set of
approximants. 
The crossover exponent $\phi$ seems to be invariant with the diffusion
rate, as already be suggested by the results of the d-log Padé
approximants.
Actually, the greatest deviations from $\phi=2$ are seen for
intermediate values $D\equiv\bar{D}/(1+\bar{D})\sim 0.5$.  
In fact, for $D=0.1$ we have $\phi=2.002\pm 0.051$ ($\phi=2.038\pm 0.201$, for
the constrained case) while for $D=0.5$, $\phi=2.134\pm 0.415$ 
($\phi=2.105\pm 0.201$, for the constrained case) and in $D=1.0$ 
we have $\phi=2.025\pm 0.090$ ($\phi=2.031\pm 0.121$).
Surprisingly, in the region close to infinite diffusion rate, or
$D=1$, the estimates of the crossover exponent show a dispersion
of the same order of the one found in the region of vanishing
diffusion $D\approx 0$. These result suggests that the crossover   
does not coincide with the mean-field result for any non-zero value of
the diffusion rate. We notice that the value $\phi=2$ is inside the
error bars for all the estimates.

\begin{figure}[h!]
\begin{center}
\vspace*{0.5cm}
\epsfig{file=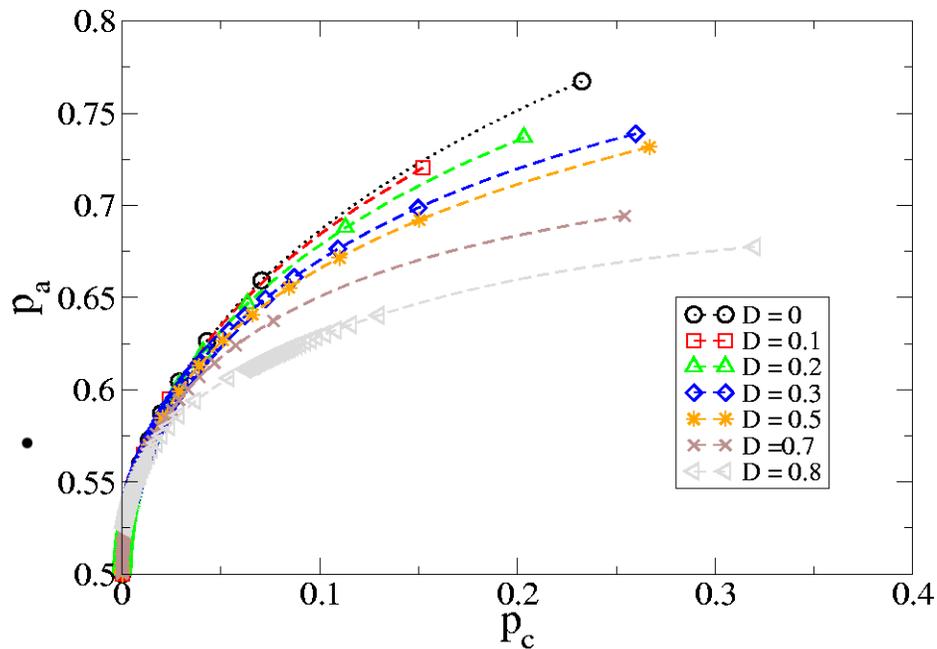,scale=0.5}
\vspace{0.4cm}
\caption{Critical lines obtained by the method of characteristics for different
  values of the diffusion rate $D$. This approach allows us to estimate
  the curves close to the multicritical point of DP-CDP crossover.} 
\label{fig6}
\end{center}  
\end{figure}

Using the method of characteristics, the equation \ref{pda} defining
the approximants may be integrated. A time-like variable $\tau$ is
introduced, and the partial differential equation will be equivalent
to a set of two ordinary differential equations:
\begin{eqnarray}
\frac{dx}{d \tau} &=& Q_M(x(\tau),y(\tau)),\nonumber\\
\frac{dy}{d\tau}  &=& R_N(x(\tau),y(\tau)).
\label{char}
\end{eqnarray}
The characteristics are the trajectories $(x(\tau),y(\tau))$ obtained
solving these equations, and the estimate of the approximant 
on each characteristic may be found through integration, once we know
the value on a particular point. It is possible to show that any
critical line which is incident on the multicritical point will itself
be a characteristic curve, and we will use this result to estimate the
DP critical line in the model. The
critical lines obtained by the method of characteristics are shown in the
figure \ref{fig6}, calculated for $D=0,0.1,0.2,0.3,0.5,0.7$, and $D=0.8$.

\begin{figure}[h!]
\begin{center}
\vspace*{0.5cm}
\epsfig{file=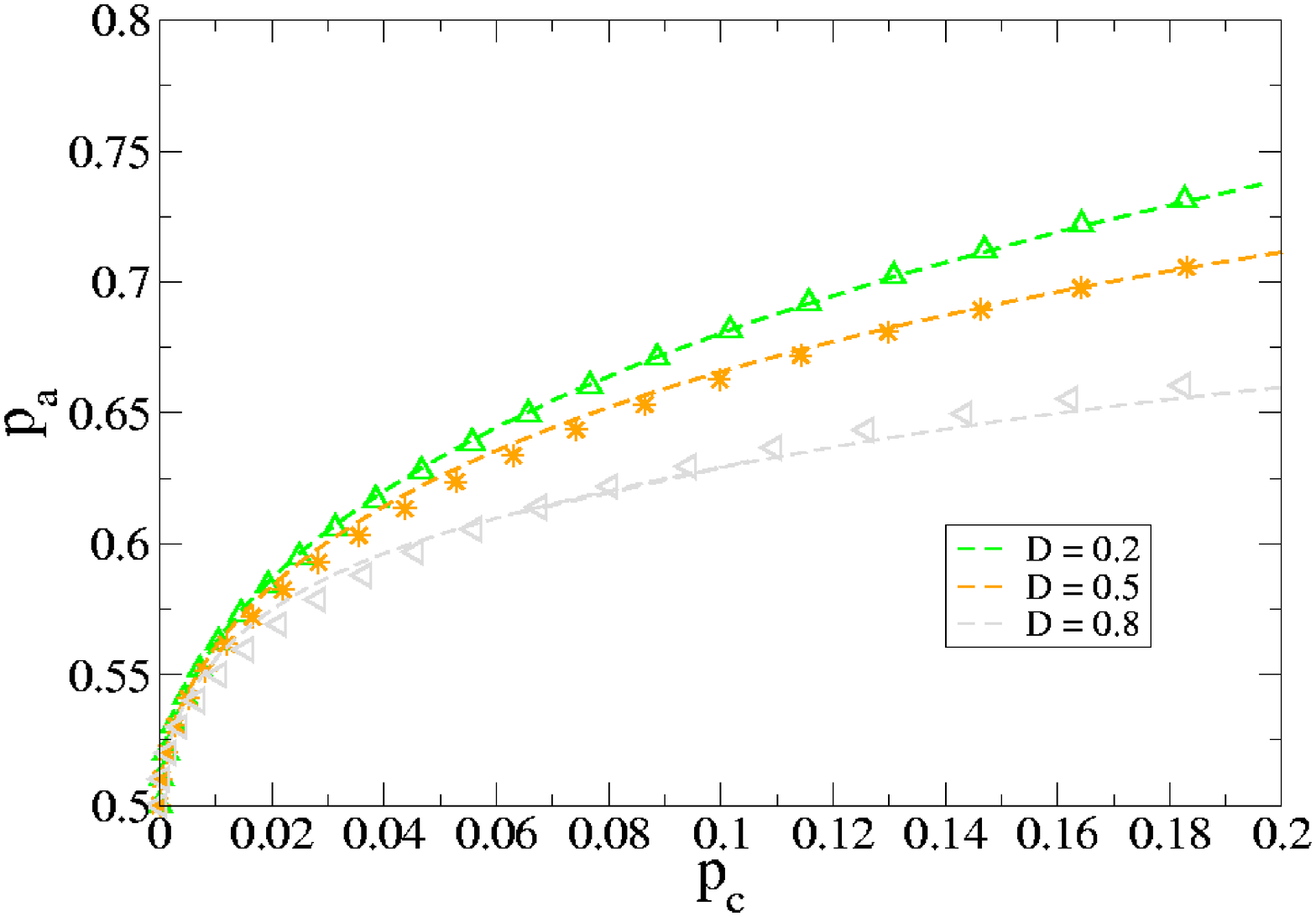,scale=0.5}\\
\vspace*{1.3cm}
\epsfig{file=./scalingcurvas.eps,scale=0.5}
\vspace{0.4cm}
\caption{Top: Estimates of critical
  lines obtained by the method of characteristics (dashed lines) and
  using the extended scaling form (symbols) are compared. Bottom:
  Critical lines calculated 
  using the extended scaling form $x(D)=z_0(D)(1-y)^2$ for different
  values of diffusion, including the infinite diffusion rate limit.}
\label{fig7}
\end{center} 
\end{figure}

Once again, the approach fails to estimate the behavior for the limit
of infinite diffusion rate.
In fact, only values to up to $D=0.8$ could be studied with reasonable
precision.  On the other hand, in \cite{ds05}, 
it is shown that a good estimate of the critical line is obtained
by the scaling function $x=z_0(1-y)^{\phi}$. A similar
procedure may be applied to the model with diffusion, adopting the
Ansatz $x=z_0(D)(1-y)^{\phi(D)}$, 
where $\phi(D)=2, \forall D$ and $z_0(D)$ is a parameter chosen
to reproduce the result in a region were we have good estimates. In the 
figure \ref{fig7} we show estimates to the critical lines from the
method of characteristics and extended scaling curves for the
diffusion rates  $D=0.2,0.5$ and $D=0.8$. In the same figure, the
extended scaling curves are depicted up to the limit of infinite
diffusion rate. In that case, the curve has a maximum, 
unlike of the behavior of the other critical lines.  Therefore, if the
hypothesis that all the critical curves can be approximated by a
scaling function such as 
$x=z_0(D)(1-y)^{\phi(D)}$ is true, then in the limit of infinite
diffusion rate 
only the limiting cases $p_b=0$ (that equals to the CP) and $p_c=0$
(voter model)  
have the same behavior that is predicted by the mean-field for this regime.  

\begin{figure}[h!]
\begin{center}
\vspace*{0.8cm}
\epsfig{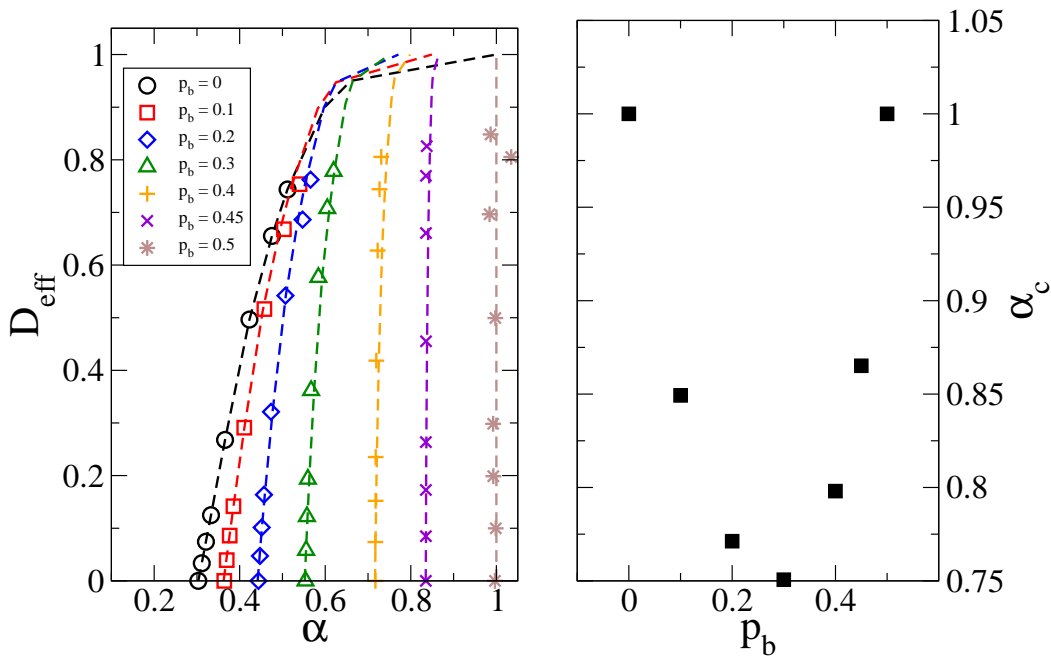}
\vspace{0.4cm}
\caption{Phase diagrams using $\alpha$ and $D_{eff}$ as variables. For each 
  value of parameter $p_b$ we have a different value for $\alpha_c$, that is
  the point where the curve achieves the infinite diffusion limit. Again
  the symbols are results from the method of characteristics and
  dashed lines are extended scaling functions.}
\label{fig8}
\end{center} 
\end{figure}

To test if the hypothesis that the critical curves can be
approximated by the function 
$x(D)=z_0(D)(1-y)^{\phi(D)}$ for all value of the diffusion rate is
reasonable, we calculate  
the behavior of $D_{eff}$ as a function of $\alpha$, in the same way
as was done
to obtain the curves shown in the figure \ref{fig2}.  Remembering
that $\alpha=(1-p_a)/p_a$ and
$D_{eff}=\alpha\tilde{D}/(1+\alpha\tilde{D})$   
we fix the value of the parameter $p_b$ and using the values of
$z_0(D)$ our results estimate the location 
the multicritical point $(\alpha_c,D_{eff}=1)$. If mean-field behavior
would be correct
for all values of $p_b$ in the infinite rate of diffusion limit, then
$\alpha_c$ should always be equal to one. 
However, as may be seen in figure \ref{fig8}, this value changes.
We also notice that in the figure \ref{fig8} an agreement
between the points obtained of the 
critical curves estimated by the approach of the characteristics and
the curves approached,  
giving support to the possibility of that $\alpha_c$ varies with $p_b$ and,
consequently with the  
parameter $p_c$, as shown in the figure \ref{fig7}.   

\begin{figure}[h!]
\begin{center}
\vspace*{1.1cm}
\epsfig{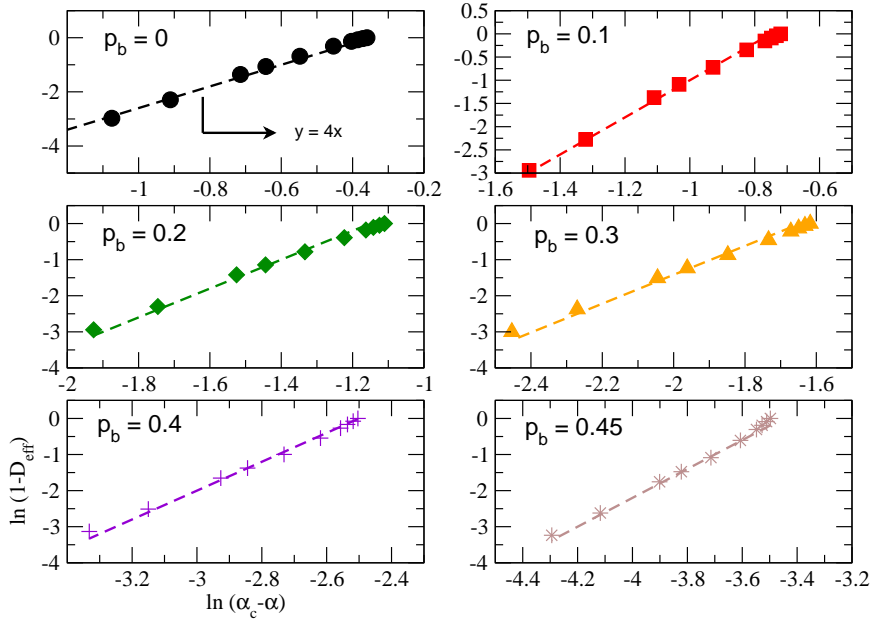}
\vspace{0.4cm}
\caption{Log-log plots for critical curves with different values of
  $p_b$, showing 
  that the choices of $\alpha_c$ from the results exhibited in figure
  \ref{fig8} result in a crossover exponent $\tilde{\phi}=4$, as
  expected for the CP.}
\label{fig9}
\end{center} 
\end{figure}

The curves for $D_{eff}$ 
as a function of $\alpha$ provide another evidence of the peculiar
behavior of the critical lines in the limit of infinite diffusion
rate. For the CP
with diffusion, at the DP-mean field change of universality class, the
estimated crossover exponent 
is $\tilde{\phi}=4$, as found in \cite{mf04}.  
In the figure \ref{fig9} we show as this value 
of the exponent $\tilde{\phi}$ is obtained using the same values for
$\alpha_c$ found in the right panel of the figure \ref{fig8}.

In summary, the results obtained by the PDA's calculated for distinct
values of $D$  
and presented in the figure \ref{fig5} are consistent with what is
known in the  
literature \cite{mf04} of the behavior expected
for the crossover
between the DP universality and the mean-field behavior. The results
presented here seem to support the conjecture that in the limit of
infinite diffusion the critical lines of the model are coincident with
the simple mean field result only in the extrema, which correspond to
the usual CP and the voter model.

\section{Conclusion}
\label{conc}
In the generalized model without diffusion, studied using series
expansions in \cite{ds05}, it was found that the DP-CDP crossover
exponent was very close to the two-site mean-field approximation
result $\phi=2$. Here we found that when diffusion is introduced in
the model, the same approximation leads to $\phi=1$ for any nonzero
finite diffusion rate and $\phi=0$ in the limit of infinite diffusion
rate. As usual, this latter result is coincident with the one which is
found applying a simple one-site mean-field approximation to the
model. However, the series analyses for the model presented here
support the conclusion that the introduction of diffusion does not
change the crossover exponent. If this conclusion remains true in the
limit of infinite diffusion, as some results above suggest, the
critical line joining the points corresponding to the CP model and the
voter model could not be the curve $p_a=1/2$ predicted by the one-site
mean field approximation, since it should be a quadratic curve in the
neighborhood of the multicritical point at $p_c=0$.

It is well known that the Pad\'e approximants will provide poor
estimates of the critical parameters in the neighborhood of a
multicritical point. Actually, this was one of the motivation for the
development of PDA's, which are suited to estimate multicritical
behavior \cite{fk77}. As expected, Pad\'e approximants show an
increasing dispersion of the estimates as the DP-CDP multicritical
point is approached, whereas PDA's lead to good estimates in this
region \cite{ds05}. As diffusion is introduced in the model, we found
out that PDA's apparently are reliable for low diffusion rates, but
as the rates are increased again the dispersion of estimates provided
by different approximants grows and for infinite diffusion rate and no
longer obtain reliable estimates from the approximants. One may
suppose that, similar to the poor performance of Pad\'e approximants
close to multicritical points, the PDA's also fail as the multicritical
point of higher order is approached for infinite diffusion rate. A
generalization of the PDA's to handle a three variable series may be
helpful to study this limit, and we are presently working in this
direction.  

Finally, in our opinion the conclusion that the crossover exponent
does not change, even in
the limit of infinite diffusion rate, which our results seem to
support, should be viewed with some caution. We stress that in order
to obtain a definite series expansion for the model, as was also
necessary in earlier studies of series expansions for similar models
with diffusion, we were forced to include the diffusive term into the part
of the evolution operator which is treated as a perturbation, and
therefore it might be possible that conclusions for high diffusion
rates are misleading. Nevertheless, for low diffusion rates we may
have more confidence in the results from the analysis of the series
expansion, and there are clear evidences that, unlike what happens in
the two-site mean field approximation, the value $\phi \approx 2$ found
in the absence of diffusion, is still valid when diffusive processes
are allowed.

\vspace{1cm}

WGD acknowledges the financial 
support from  Funda\c{c}\~ao de Amparo \`a Pesquisa do
Estado de S\~ao Paulo (FAPESP) under Grant No. 05/04459-1.
JFS is grateful for the support provided by
project PRONEX-CNPq-FAPERJ/171.168-2003.

\end{document}